\documentclass{elsart}

\usepackage{graphicx,amssymb}

\begin{document}
\begin{frontmatter}

\title{Extragalactic UHE proton spectrum and prediction for iron-nuclei 
flux at $10^8 - 10^9$~GeV}

\author[GS,INR]{V. S. Berezinsky},
%\ead{veniamin.berezinsky@lngs.infn.it}
\author[INR]{S. I. Grigorieva},
%\ead{grigorieva@inr.npd.ac.ru}
\author[AOK]{B. I. Hnatyk}
%\ead{hnatyk@kiev.ua}

\address[GS]{INFN, Laboratori Nazionali del Gran Sasso, I-67010 Assergi
(AQ), Italy}
\address[INR]{Institute for Nuclear Research, 60th October Revolution prospect
7A, 117312 Moscow, Russia}
\address[AOK]{Astronomical Observatory of Kiev National
University, 3 Observatorna street, 04053 Kiev, Ukraine}

\begin{abstract}
We investigate the problem of transition from galactic cosmic rays
to extragalactic ultra high energy cosmic rays. Using the model
for extragalactic ultra high energy cosmic rays and observed
all-particle cosmic ray spectrum, we calculate the galactic
spectrum of iron nuclei in the energy range $10^8 - 10^9$~GeV. The
flux and spectrum predicted at lower energies agree well with the
KASCADE data. The transition from galactic to extragalactic cosmic
rays is distinctly seen in spectra of protons and iron nuclei,
when they are measured separately. The shape of the predicted iron
spectrum agrees with the Hall diffusion.

\end{abstract}

\begin{keyword} cosmic rays \sep ultra-high energy cosmic rays \sep 
knee

\PACS 96.40 \sep 96.40.De \sep 98.70.Sa
\end{keyword}
\end{frontmatter}

%%%%%%%%%%%%%%%%%%%%%%%%%%%%%%%%%%%%%%%%%%%%%%%%%%%%%%%%%%%%%%%%%%%%
\section{Introduction}
\begin{figure}[t]
\centerline{\includegraphics[width=0.7\textwidth,height=0.7\textwidth]
{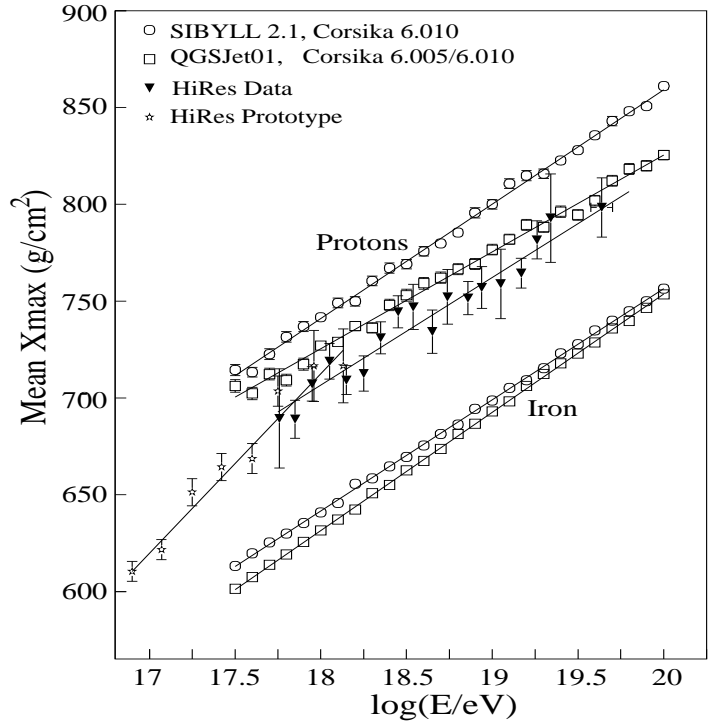}}
%%%%%%%,angle=270]{phases.eps}}\cite
%\vskip-5mm
\caption{\em The HiRes data \cite{Sokol} on mass composition
(preliminary). The measured $x_{\rm max}$ at $E\geq 1\times
10^{18}$~eV are in a good agreement with QGSJet-Corsika prediction
for protons. \label{hires}}
\end{figure}

\begin{figure}[t]
\centerline{\includegraphics[width=0.9\textwidth]{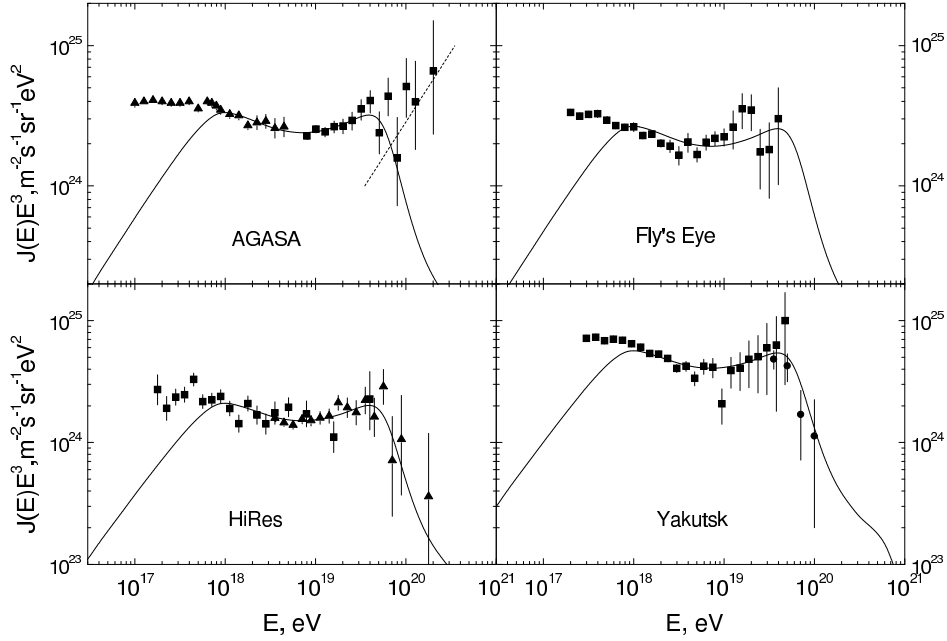}}
%%%%%%%,angle=270]{phases.eps}}\cite
%\vskip-5mm
\caption{\em
Comparison of calculated spectra for non-evolutionary model
(see the text) with observational data. There is a good agreement of
a dip centered at $E \sim 1\times 10^{19}$~{\rm eV} with all data. The dip
is produced due to pair-production $p+\gamma_{\rm CMB}\to p+e^{+}+e^{-}$
on CMB radiation. AGASA excess needs for its explanation another
component of UHECR (shown by dashed curve), which can be due to
one of the top-down scenarios \cite{ABK}.
\label{dip}}
\end{figure}

There are now convincing evidences that ultra high energy cosmic
rays (UHECR) in the energy range $1\times 10^{18} - 8\times
10^{19}$~eV are extragalactic protons. These evidences include:
(i) Measurement by HiRes detector \cite{Sokol} of $x_{\rm max}$ in
EAS longitudinal development favours protons as primaries at $E
\geq 1\times 10^{18}$~eV (see Fig.~\ref{hires}), (ii) The energy
spectra measured by Akeno-AGASA \cite{AGASA}, Fly's Eye \cite{FE},
HiRes \cite{HiRes} and Yakutsk \cite{Yakutsk} detectors clearly
show  the {\em dip} \cite{BGG} (see Fig.\ref{dip})  , which is a
signature of interaction of extragalactic protons with CMB
radiation, (iii) The beginning of the GZK cutoff up to $E \approx
8\times 10^{19}$~eV is seen in all data, including that of AGASA
(see Fig.~\ref{dip}).

It is interesting to note that the excess, detected by AGASA \cite{AGASA-anis}
at $E\sim 10^{18}$~eV from directions to the galactic sources,
Galactic Center and Cygnus, may be  naturally interpreted as
evidence of diffuse {\em extragalactic} flux at $E\geq 1\times 10^{18}$~eV.
Indeed, if transition from galactic to extragalactic diffuse flux
occurs due to failure of magnetic confinement in the Galaxy, the
unconfined flux from the galactic sources should become visible above
energy of the transition.

On the other hand, not all experimental data agree with pure proton
composition at $E \geq 1\times 10^{18}$~eV. While data of
HiRes-MIA\cite{casa-mia}
and Yakutsk \cite{yakutsk1} support  such
composition, the data of other detectors, such as Fly's Eye \cite{FE1},
Haverah Park \cite{HP} and Akeno \cite{akeno} favour the mixed
composition at $E \geq 1\times 10^{18}$~eV. The indirect
confirmation of the proton composition at $E \geq 1\times 10^{18}$~eV
comes from the KASCADE data \cite{kascade}, which show disappearance of the
heavy nuclei from the CR flux at much lower energies.

UHECR have a problem with the highest energy events.
First of all, 11 AGASA events at $E\geq 1\times 10^{20}$~eV comprise
significant excess (not seen by HiRes) over predicted flux
(see Fig.~\ref{dip}). This excess may be interpreted as the new
component of UHECR, probably due to one of the top-down
scenarios (see \cite{ABK}). Even if
to exclude the AGASA data from analysis, the problem with particles
at $E>1\times 10^{20}$~eV remains. There are three events with
$E>1\times 10^{20}$~eV observed by other detectors, namely, one Fly's
Eye event with $E \approx 3\times 10^{20}$~eV, one HiRes event with
$E \approx 1.8\times 10^{20}$~eV and one Yakutsk event with
$E\approx 1.0 \times 10^{20}$~eV. The attenuation length of protons
with energies $(2 - 3)\times 10^{20}$~eV is only 20 - 30~Mpc. With
correlations of UHECR directions with AGN (BL Lacs) observed
at $(4-8)\times 10^{19}$~eV \cite{TT}, the AGN must be observed in the
direction of these particles. If to exclude the correlations with AGN
from analysis, the propagation in the strong magnetic fields becomes
feasible \cite{strongH}. Nevertheless, also in this case the lack
of the nearby sources in the direction of highest energy events (e. g.
at $E \sim 3\times 10^{20}$~eV) remains a problem for reasonable
local field $H \sim 1$~nG. Indeed, the deflection angle
$\theta \sim l_{\rm att}/r_H= 3.7^{\circ}H_{\rm nG}$
is small, and the source should be seen within this angle.

\begin{figure}[t]
%\begin{tabular}{c}
%\centerline{\includegraphics[width=0.8\textwidth]{kascade38.eps}}
%\\
\centerline{\includegraphics[width=0.8\textwidth]{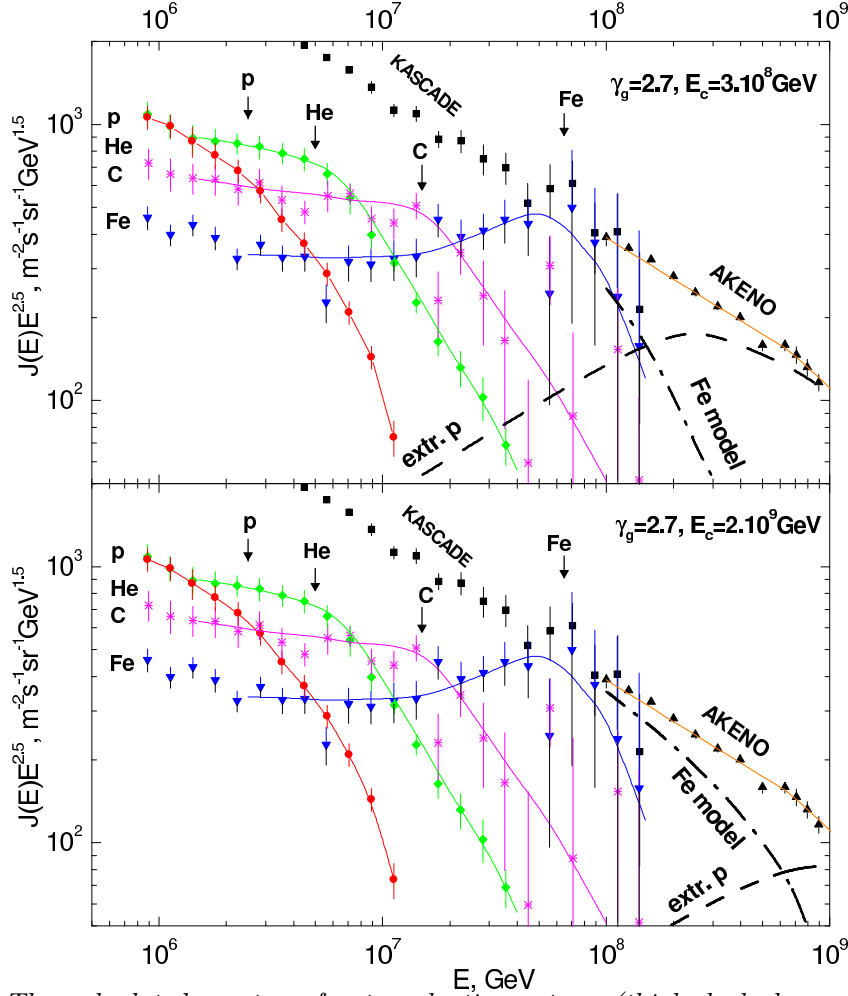}}
%\end{tabular}
\vskip-5mm 
\caption{\em The calculated spectra of extragalactic
protons (thick dashed curve labeled ``extr. p'') and of galactic
iron nuclei (dash-dotted thick curve labeled ``Fe model'') in
comparison with the Akeno and KASCADE data. The values of
$\gamma_g$ and $E_c$ used in the calculations are indicated  in
each panel. The all-particle spectrum of Akeno is shown by filled
triangles and by solid curve. The sum of the curves ``extr. p''
and ``Fe model'' is exactly equal to  all-particle spectrum of
Akeno, given by curve ``AKENO''. The data of KASCADE for different
nuclei are shown as : protons --  by filled circles, helium --  by
diamonds, carbon -- by stars, and iron -- by inverted triangles.
The thin full curves smoothly connect the data points. The arrows
labeled by p, He, C and Fe show the positions of corresponding
knees, calculated as $E_Z=ZE_p$, with $E_p=2.5 \times 10^{6}$~GeV.
One can notice the satisfactory agreement between calculated and
observed positions of the knees. \label{kascade}}
\end{figure}

Another problem with UHECR exists at its low energy edge.
The transition from galactic to extragalactic cosmic rays may occur at
position of the second knee.  The energy of the second knee varies from
$4\times 10^{17}$~eV to $8\times 10^{17}$~eV in different experiments
(Fly's Eye -- $4\times 10^{17}$~eV  \cite{FE}, Akeno --
$6\times 10^{17}$~eV \cite{Akeno2},
HiRes -- $7\times 10^{17} $~eV \cite{HiRes} and Yakutsk --
$8\times 10^{17}$~eV \cite{Yakutsk}).
This energy is close to $E\sim 1\times 10^{18}$~eV, where
according to the HiRes data \cite{casa-mia} protons start to dominate
the flux.
On the other hand, KASCADE data \cite{kascade} show disappearance of
galactic cosmic rays at much lower energy. In Fig.~\ref{kascade} one
can see that positions of the proton, helium, carbon and iron knees
can be tentatively accepted as $2.5\times 10^{15}$~eV,~
$5.5\times 10^{15}$~eV,~ $1.8\times 10^{16}$~eV,~
and $7.0\times 10^{16}$~eV, respectively (see arrows in Fig.~\ref{kascade}).

How the gap between the knee for iron nuclei and the beginning of the
extragalactic component is filled? Why the transition reveals itself
in the form of hardly noticeable feature in all-particle spectrum?

%%%%%%%%%%%%%%%%%%%%%%%%%%%%%%%%%%%%%%%%%%%%%%%%%%%%%%%%
\section{The model}
\label{model}
We will construct here the phenomenological model, which predicts the
spectrum of iron nuclei in the energy range
$8\times 10^7 - 1\times 10^9$~GeV. Following Ref.~\cite{BGG},
we will calculate the spectrum of extragalactic protons in the energy
range $10^8 - 10^{12}$~GeV. Subtracting this proton spectrum from the
observed all-particle spectrum (we shall use Akeno-AGASA data),
we obtain residual spectrum, which, inspired by the KASCADE data,
we assume to be comprised by iron nuclei.
This spectrum will be compared with that measured by KASCADE. The
justification of this model consists in confirmation of the dip in
the calculated proton spectrum by observations. One of the results
is description
of transition from galactic to extragalactic cosmic rays. The predicted
physical quantity is the flux of iron nuclei and their spectrum at
$10^8 - 10^9$~GeV.
%%%%%%%%%%%%%%%%%%%%%%%%%%%%%%%%%%%%%%%%%%%%%%%%%%%%%%%%%
\subsection{Spectrum of extragalactic protons}
We will calculate the extragalactic proton spectrum, following Ref.~\cite{BGG},
in the model with the following assumptions.

We assume the generation spectrum of a source
\begin{equation}
Q_g(E_g,z)=\frac{L_p(z)}{\ln\frac{E_c}{E_{\rm min}}+\frac{1}{\gamma_g-2}}
q_{\rm gen}(E_g),
\label{gen}
\end{equation}
with
\begin{equation}
q_{\rm gen}(E_g)=\left\{ \begin{array}{ll}
1/E_g^2                      ~ &{\rm at}~~ E_g \leq E_c\\
E_c^{-2}(E_g/E_c)^{-\gamma_g}~ &{\rm at}~~ E_g \geq E_c
\end{array}
\right.
\label{q-gen}
\end{equation}

The diffuse spectrum is calculated as
\begin{equation}
J_p(E)=\frac{c}{4\pi}~\frac{{\mathcal L}_0 }
{\ln\frac{E_c}{E_{min}}+\frac{1}{\gamma_g-2}} \int_0^{z_{max}} dt
q_{\rm gen}\left(E_g(E,z),E\right) \frac{dE_g}{dE}, \label{diff}
\end{equation}
where ${\mathcal L}_0=n_sL_p$ is emissivity (with $n_p$ and $L_p$
being the comoving  density of the sources  and luminosity,
respectively), $E_g(E,z)$ is generation energy of a proton at
epoch $z$, $dE_g/dE$ is given in Refs.~\cite{BG88,BGG1} as
\begin{equation}
\label{dE_g/dE}
\frac{d E_g(z_g)}{dE} = (1+z_g) exp \left\{\frac{1}{H_0} \int_0^{z_g} dz
\frac{(1+z)^2}{\sqrt{\Omega_m(1+z)^3+\Omega_{\Lambda}}} \left (
\frac{db_0(E')}{dE'}\right )_{E'=(1+z)E_g(E,z)} \right \},
\end{equation}
where $b_0(E)$ is the proton energy loss $dE/dt$ at $z=0$,
and $dt$ is given as
\begin{equation}
dt=\frac{dz}{H_0(1+z)\sqrt{\Omega_m(1+z)^3+\Omega_{\Lambda}}},
\label{dt/dz}
\end{equation}
with $H_0, \Omega_m,$ and $\Omega_{\Lambda}$ being the Hubble
constant, relative cosmological density of matter and that of vacuum energy,
respectively.

In calculation of the flux  according to Eq.~(\ref{diff}) we use
the non-evolutionary model, taking ${\mathcal L}_0$ not dependent
on redshift $z$. Since we are interested in energies $E \geq
1\times 10^8$~GeV the maximum redshift appears in Eq.~(\ref{diff})
automatically due to sharp increase of $E_g(z)$ at large $z$. The
control calculations with $z_{\rm max}=4$ give the the identical
results. Notice that in evolutionary models the calculated flux is
sensitive to $z_{\rm max}$.

The calculated spectrum is displayed in Fig.~\ref{dip} with
$E_c=1\times 10^{9}$~GeV and $\gamma_g=2.7$ in comparison
with observational data. In the further calculations we fix $\gamma_g=2.7$
as the best fit to the observational data displayed in Fig.~\ref{dip} and
consider $E_c$ as a free parameter.

%%%%%%%%%%%%%%%%%%%%%%%%%%%%%%%%%%%%%%%%%%%%%%%%%%%%%%%%%%%%%%%%%%%%%
\subsection{Galactic iron spectrum and transition from galactic to
extragalactic cosmic rays}
\begin{figure}[t]
\centerline{\includegraphics[width=0.7\textwidth]{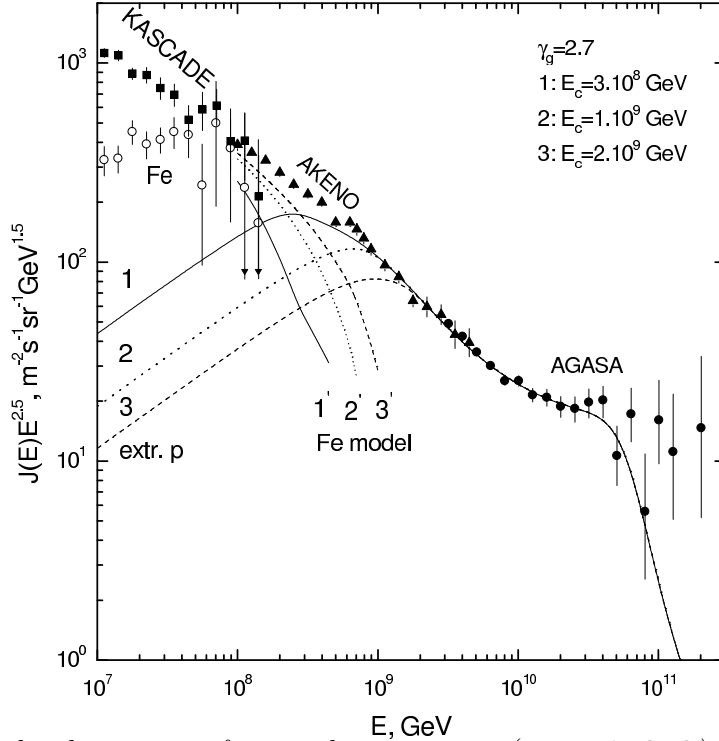}}
%%%%%%%%,angle=270]{phases.eps}}
\vskip-5mm
\caption{\em Calculated spectrum of extragalactic protons (curves $1,~2,~3$)
and of galactic iron spectra (curves $1', 2', 3'$) compared with
all-particle spectrum from Akeno and AGASA experiments. The galactic
iron spectrum is obtained by subtraction of the calculated proton
spectrum from the all-particle spectrum. The pairs of curves $1$ and $1'$,
$2$ and  $2'$, $3$ and $3'$
correspond to $E_c$ equal to $3\times 10^8$~GeV, $1\times 10^9$~GeV,
and $2\times 10^9$~GeV, respectively. The intersections of the curves
$1-1'$, $2-2'$ and $3-3'$ give the transition from galactic
(iron) to extragalactic (protons) components, which occurs at
$1.5\times 10^8$~GeV,~ $3.6\times 10^8$~GeV and  $6.2 \times 10^8$~GeV,
respectively. The KASCADE data are shown by filled squares for 
all-particle fluxes and by open circles  - for iron nuclei fluxes.
\label{trans}}
\end{figure}

\begin{figure}[t]
\centerline{\includegraphics[width=0.7\textwidth]{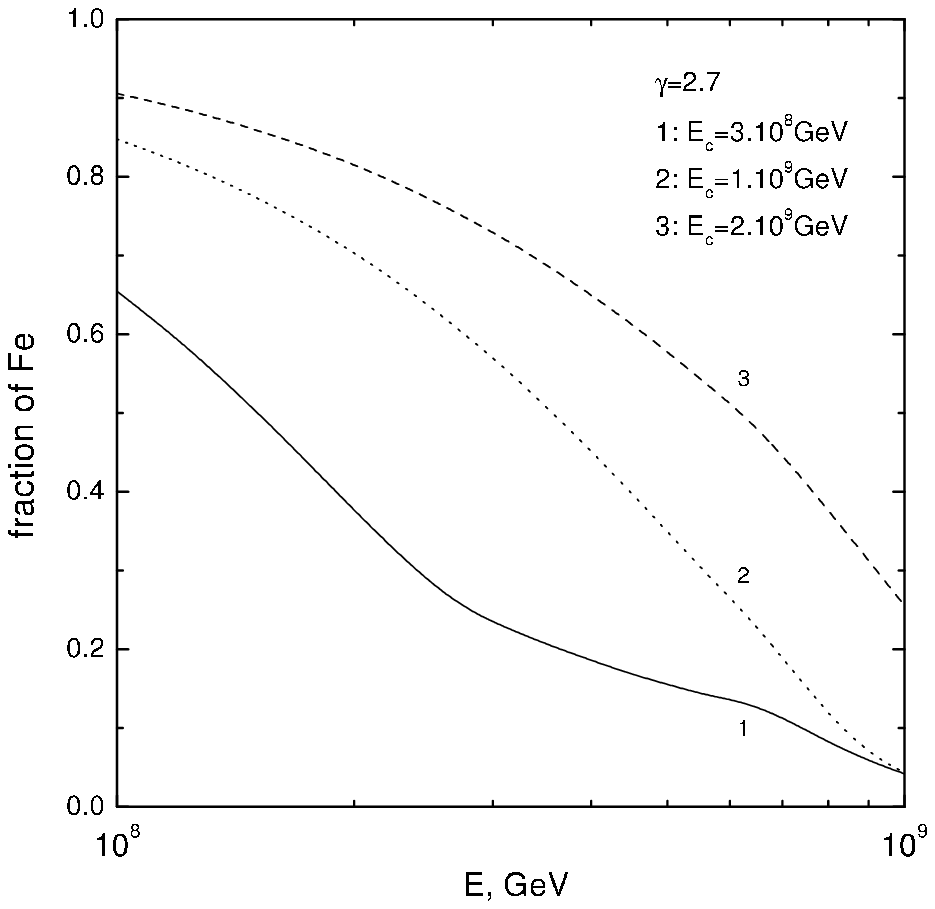}}
%%%%%%%%,angle=270]{phases.eps}}
\vskip-5mm
\caption{\em  Fraction of iron nuclei in the total flux as function of energy.
\label{fraction}}
\end{figure}

In Fig.~\ref{trans} we show the iron galactic spectrum found as
subtraction of the extragalactic proton spectrum calculated above,
from all-particle spectrum (Akeno-AGASA). The transition from galactic
to extragalactic cosmic rays occurs at crossing of proton and iron
spectra, which is very noticeable and occurs at energies
$1.5\times 10^8$~GeV, $3.6 \times 10^8$~GeV and $6.2 \times 10^8$~GeV
for the curves 1-$1'$,~ 2 - $2'$~ and 3 - $3'$, respectively.
In Fig.~\ref{fraction} the fraction of iron nuclei in the total flux is
shown as function of energy. Notice, that in our model
in energy range $(1 - 10)\times 10^8$~GeV the total flux is comprised only
by protons and iron. The Haverah Park data \cite{HP} confirm that it
can be the case. The analysis of the Haverah Park data has been performed
for energy-independent composition, however comparison of the obtained
composition (34\% of protons and  66\% of iron) with our average value favours
large $E_c$.

The flux of iron nuclei calculated here fit well the KASCADE data at energy
$E\sim 10^8$~GeV and below  (see Fig.~\ref{kascade}).

Why there is no pronounced feature in the total spectrum which
corresponds to transition from galactic to extragalactic component?
Such feature is inevitably faint when both components have
approximately equal spectrum exponents. This is our case: the Akeno
spectrum up to $7\times 10^7$~GeV is characterised by $\gamma =3.0$,
while at higher energies - by $\gamma=3.25$. However,
one can see from Fig.~\ref{kascade} that intersection of extragalactic
proton spectrum (``extr. p'') and galactic iron spectrum (curve ``Fe-model'')
is characterised by quite prominent feature. Therefore, measuring
the spectra of iron and protons separately, one can observe the
transition quite distinctively.

\subsection{Visibility of galactic sources at $E \gtrsim 1\times 10^{9}$~GeV}

We argued above that at energy $E > E_c \sim 1\times 10^{9}$~GeV the
extragalactic flux dominates.  However, as the AGASA observations show 
\cite{AGASA-anis} at these energies the excess of events in the direction
of some galactic sources is observed. We shall show in
this section that visibility of galactic sources at $E > E_c$ is a
natural prediction of our model.   

If galactic sources accelerate particles to energies higher than 
$1\times 10^9$~GeV their ``direct'' flux can be seen, while the
produced diffuse flux should be small because of short confinement
time in the galaxy. If generation spectrum is dominated by protons,
the ``direct''flux must be seen as the protons, while the diffuse galactic
flux is presented by the heaviest nuclei. 

We shall argue here that galactic point sources should be observed as the
extensive sources, as it occurs in the AGASA observations \cite{AGASA-anis}.

Propagation of the protons from the galactic sources takes place in
quasi-diffusive regime with the large diffusion coefficient. The
protons arrive at an observer along a trajectory in the {\em regular} 
galactic magnetic field, while the small-angle scattering in the
random magnetic field provides the angular distribution of protons in
respect with the main direction along the trajectory. 
Because of the large proton energy, the scattering occurs
mainly on the basic scale of the turbulent (random) magnetic field 
$l_c \sim 100$~pc. The scattering angle $\theta$ corresponds to 
$\langle \theta^2\rangle=l_c^2/r_L^2$, where $r_L$ is the Larmor
radius in the basic magnetic field. Using 
$\langle B_{\perp}^2\rangle=(2/3)\langle B^2 \rangle$ for the field 
perpendicular to a particle trajectory  one obtains 
\begin{equation}
\langle \theta^2 \rangle=\frac{2}{3}e^2 l_c^2 \langle B^2 \rangle/E^2,
\label{sc-angle}
\end{equation}
where $e$ is electric charge of the proton. 

After $n$ scattering in the random magnetic field the angle with
respect to the direction of the regular trajectory 
becomes \cite{Alcock,Roulet} ~~
$\langle \phi^2 \rangle = \frac{1}{3} n \langle \theta^2 \rangle$ .

Using $n=r/l_c$, where $r$ is a distance to the source along the
regular trajectory, one finally obtains
\begin{equation}
\langle \phi^2 \rangle = \frac{2}{9}e^2 r l_c \langle B^2 \rangle/E^2 ,
\label{deflection}
\end{equation}
similar to formulae used in Ref. \cite{Waxman,Sigl}. 

The distribution of arriving particles over the angles is the normal
Gaussian, with dispersion $\sigma=\sqrt{\langle \phi^2 \rangle} $.
For a distance $ r=10$~kpc,  $B=1~\mu$G and 
$E= 1\times 10^9$~GeV one obtains $\sigma =25^{\circ}$ in a  good
agreement with the AGASA observations. 

At larger energies and smaller distances the sources are seen as the 
point-like ones.
%%%%%%%%%%%%%%%%%%%%%%%%%%%%%%%%%%%%%%%%%%%%%%%%%%%%%%%%%%%%%%%
\section{Discussion and conclusions}
One of the assumptions of our model is that at high energies
$E> 1\times 10^8$~GeV only iron nuclei as galactic component
survive. This assumption is inspired first of all by the KASCADE data
(see Fig.~\ref{kascade}) which show how light nuclei gradually
disappear from the spectrum with increasing the energy. This result
is in the accordance with other
experiments, e.g. \cite{AMANDA}, which also demonstrate that the mean atomic weight
$\langle A \rangle$ (and hence $\langle Z \rangle$) increases with
energy.

The good agreement of calculated iron spectrum with that measured
by KASCADE at $E\leq 10^8$~GeV might be delusive. The published spectra
are still preliminary and they are slightly different in different
publications. In effect, it is more proper to  speak only about
rough agreement of
calculated spectrum with one measured by KASCADE. On the other hand,
we have a free parameter $E_c$ with help of which we can fit our
spectrum to the KASCADE data. Such fit would restrict our predictions
for the Fe/p ratio at higher energies (see Fig.~\ref{fraction}).

In principle there are three classes of models explaining the knee
(for a review and references see \cite{Hoerandel}):\\
(i) Rigidity-dependent exit of cosmic rays from the Galaxy, e. g.
\cite{Ptuskin,Lagutin,Roulet1}.\\
(ii) Rigidity-dependent acceleration \cite{Biermann,Berezhko,Erlykin}.\\
(iii) Interaction-produced knee \cite{Tkachyk,Roulet2}.\\
For more references see \cite{Hoerandel}.

As far as the spectra are concerned, the first two classes can result
in identical conclusions. For example the spectra of nuclei in
Ref.~\cite{Biermann} with acceleration in presupernova winds have the
same rigidity dependence as in the models of class (i).  We shall
restrict ourselves in further discussion by models of class (i).

The data of KASCADE (Fig.~\ref{kascade}) are compatible with rigidity
dependent knees for different nuclei, as it must be in models (i) and
as it might be in models (ii).

The positions of the proton, helium, carbon and iron knees in the
KASCADE data can be
tentatively accepted as $2.5\times 10^6$~GeV,~ $5.5\times 10^6$~GeV,~
$1.8\times 10^7$~GeV,~
and $7.0\times 10^7$~GeV, respectively (see arrows in Fig.~\ref{kascade}).
The accuracy of the knee positions is not good enough and is
different for different knees. While position of the proton knee,
$E_p \approx 2.5\times 10^6$~GeV, agrees with many other measurements
(see \cite{Hoerandel} for a review), the position (and even existence)
of the iron knee is uncertain.
The energies given  above should be considered not more than
indication. The future
Kascade-Grande\cite{kascade-g} data will clarify the situation with
the iron knee. However, it is interesting to note that the positions
of the nuclei knees, taken above from the KASCADE data, coincide well with
simple rigidity model of particle propagation in the Galaxy. In this
model $E_Z=ZE_p$, where Z is a charge of a nucleus. Taking
$E_p=2.5\times 10^{6}$~GeV, one obtains for helium, carbon and iron knees
$E_{\rm He}=5.0\times 10^6$~GeV,~ $E_{\rm C}=1.5\times 10^7$~GeV and
$E_{\rm Fe}=6.5\times 10^7$~GeV, respectively, in a good agreement
with the KASCADE data cited above (see Fig. 3).

In the rigidity-dependent models (i) one can predict the spectrum
shape below and above the knee.
{\em Below} the knee, in the case of the Kolmogorov spectrum of random
magnetic fields  the diffusion coefficient $D(E) \propto E^{1/3}$,
and the generation index for galactic cosmic rays is $\gamma_g = 2.7-0.33
\approx 2.35$. If {\em above} the knee $D(E) \propto E^k$, the spectral
index $\gamma= \gamma_g+k$.  At extremely high energy, when the Larmor
radius $r_L$ becomes much larger than the basic scale of magnetic
field coherence $l_c$, the diffusion reaches the asymptotic regime with
$D(E) \propto E^2$. In this case $\gamma=4.35$. The critical energy
is $E_{cr}=ZeB_0l_c =2.4\times 10^{9}$~GeV for iron (Z=26), where
$l_c= 100$~pc and the regular magnetic field on the basic scale is
$B_0 \sim 1\mu$G. Therefore, the asymptotic regime with $\gamma=4.35$
is valid at $E \gg 2.4\times 10^{9}$~GeV and in the energy region of
interest the intermediate regime with non-power law spectrum may
occur.

Below the knee the {\em longitudinal}
diffusion (along the regular field $\vec{B}_0$) dominates.
It is provided by condition  of strong magnetizing. At higher energy
the role of {\em transverse} diffusion might be important.
An example of it can be given by the {\em Hall diffusion}
\cite{Ptuskin,Roulet}, associated with the drift of particles
across the regular magnetic field. In this case above the knee
$E>ZE_p$,~ ~$D(E)\propto E$ and $\gamma= 3.35$.

In our model the iron spectra are different for the cases
$1'$~ ($E_c=3\times 10^8$~ GeV),~$2'$~ ($E_c=1\times 10^9$~ GeV) and
$3'$~ ($E_c=2\times 10^9$~ GeV) (see Fig.~\ref{trans}). In fact all
spectra are not power-law, but in the effective power-law
approximation they can be roughly characterised by  $\gamma \approx 3.9$
for spectrum $1'$,~ $\gamma \approx 3.4$~ for spectrum $2'$ and
$\gamma= 3.3$ for spectrum $3'$. These spectra, especially $2'$ and $3'$
are  consistent with the Hall diffusion, which predicts
$\gamma=3.35$.

If galactic sources accelerate particles to energy higher than 
$1\times 10^{9}$~GeV, they should be observed in extragalactic
background by the ``direct'' flux. Due to multiple scattering of
protons in the galactic magnetic field the sources look as extensive
ones with the typical angular size $\sim 20^{\circ}$ at distance 
$\sim 10$~kpc.

In conclusion, the model for UHE proton propagation (Section~\ref{model})
combined with measured all-particle spectrum (taken as Akeno-AGASA data)
predicts the galactic iron spectrum in energy range $1\times 10^8 -
1\times 10^9$~GeV. The predicted flux agrees well at
$E \sim 8\times 10^7$~GeV and below with the KASCADE data. The transition from
galactic to extragalactic cosmic rays occurs in the energy region
of the second knee and is distinctly seen if iron and proton spectra
are measured separately. In all-particle spectrum  this transition
is characterised by a faint feature because spectral indices of galactic 
component at $E < 1 \times 10^9$~GeV and extragalactic component 
at $E > 1 \times 10^9$~GeV are close to each other ($\Delta\gamma =0.25$).
The spectrum of iron nuclei in $1\times 10^8 - 1\times 10^9$~GeV
energy range agrees with the Hall diffusion.

\section{Acknowledgments}

We thank transnational access to research infrastructures (TARI)
program through the LNGS TARI grant contract HPRI-CT-2001-00149. 
This work is partially supported  by 
Russian grants  RFBR 03-02-16436a, RFBR 04-02-16757 and LSS - 1782.2003.2.\\
We thank V. Ptuskin and our collaborators in TARI project R. Aloisio
and A. Gazizov for valuable discussions.


\footnotesize
\frenchspacing

\begin{thebibliography}{99}

\bibitem{Sokol} P. Sokolsky, ``The High Resolution Fly's Eye - Status
and Preliminary Results on Cosmic Ray Composition above $10^{18}$~eV'',
Proc. of SPIE Conf on Instrumentation for Particle Astrophysics,
Hawaii, (2002);

G. Archbold and  P. V. Sokolsky (for the HiRes Collaboration),
Proc. of 28th International Cosmic Ray Conference, 405 (2003).

\bibitem{AGASA}
M. Takeda et al (AGASA collaboration), Astroparticle Phys. {\bf 19},
447 (2003).

\bibitem{FE} D. J. Bird et al, Ap.J. {\bf 441}, 144 (1995).

D. J. Bird et al (Fly's Eye collaboration), Ap. J., {\bf 424}, 491 (1994).

\bibitem{HiRes} T. Abu-Zayyad et al (HiRes collaboration)\ 2002a,
astro-ph/0208243;\\
T. Abu-Zayyad et al (HiRes collaboration)\ 2002b, astro-ph/0208301.

\bibitem{Yakutsk} A. V. Glushkov et al (Yakutsk collaboration),
JETP Lett. {\bf 71}, 97 (2000);\\
A. V.  Glushkov and M. I. Pravdin, JETP Lett. {\bf 73}, 115 (2001).

\bibitem{BGG} V. S. Berezinsky, A. Z. Gazizov, and S. I. Grigorieva,
Proc. of Int. Workshop "Extremely High Energy
Cosmic Rays" (eds M. Teshima  and T. Ebisuzaki) Universal Academy
Press Inc., Tokyo, Japan, 63 (2003), astro-ph/0302483.

\bibitem{AGASA-anis}  N. Hayashida et al (AGASA collaboration), 
Astroparticle Physics {\bf 10}, 303 (1999).

\bibitem{casa-mia} T. Abu-Zayyad et al, Astrophys. J. {\bf 557}, 686 (2001);\\
T. Abu-Zayyad et al, Phys. Rev. Lett., {\bf 84}, 4276 (2000).

\bibitem{yakutsk1} A. V.  Glushkov et al (Yakutsk collaboration),
JETP Lett. {\bf 71}, 97 (2000).

\bibitem{FE1} D. Bird et al, Phys Rev. Lett.,{\bf 71}, 4276 (1993).

\bibitem{HP} M. Ave et al, Astroparticle Phys., {\bf 19}, 61 (2003).

\bibitem{akeno}  K. Shinozaki et al, Proc. of 28th International
Cosmic Ray Conference (ICRC), {\bf 1}, 401 (2003).

K. Shinozaki et al, Ap.J. {\bf 571}, L117 (2002);\\
T. Doi et al, Proc. 24th ICRC (Rome) {\bf 2}, 685 (1995).

\bibitem{kascade}
%" The Physics of the Knee in the Cosmic Ray Spectrum"
K.-H. Kampert et al (KASCADE-Collaboration),  Proceedings of 27th ICRC,
volume "Invited, Rapporteur, and Highlight papers of ICRC", 240 (2001);\\
J. R. Hoerandel et al (KASCADE Collaboration), Nucl. Phys. B
(Proc. Suppl.) {\bf 110}, 453 (2002).

\bibitem{ABK}
R. Aloisio, V. Berezinsky, M.Kachelriess, hep-ph/0307279.

\bibitem{TT}
P. G. Tinyakov and I. I. Tkachev, JETP Lett., {\bf 74}, 445 (2001).

\bibitem{strongH} G. Sigl, M. Lemoine, P. Biermann, Astroparticle Phys.,
{\bf 10}, 141 (1999);\\
D. Harari, S. Mollerach, E. Roulet, JHEP 0207, 006 (2002);\\
H. Yoshiguchi  et al, Ap.J., {\bf 586}, 1211 (2003).

\bibitem{Akeno2} M. Nagano et al, J. Phys. G. Nucl. Part. Phys.,
{\bf 18}, 423 (1992).


\bibitem{BG88} V. S. Berezinsky and S. I. Grigorieva,
Astron. Astroph., {\bf 199}, 1 (1988).

\bibitem{BGG1} V. S. Berezinsky, A. Z. Gazizov, and S. I. Grigorieva,
2002a, hep-ph/0204357;\\
V. S. Berezinsky, A. Z. Gazizov, and S. I. Grigorieva, 2002b, astro-ph/0210095.

\bibitem{AMANDA} K. Rawlins (for SPASE and AMANDA collaborations),
Proc. of 28th International Cosmic Ray Conference, 173 (2003).
% "Measurement of the Cosmic Ray Composition at the Knee with the SPASE-2/AMANDA-B10 Detectors"

%\bibitem{agasa-s}
%N.Hayashida et al (AGASA collaboration), Astroparticle  Physics {\bf 10},
%303 (1999).

\bibitem{Alcock}
C. Alcock and S. Hatchett, Ap.J., {\bf 222}, 456 (1978).

\bibitem{Roulet}
D. Harari, S. Mollerach, E. Roulet and F. Sanchez, JHEP 0203, 045 (2002).

\bibitem{Waxman}
E.Waxman and J. Miralda-Escude, Ap.J., {\bf 472}, L 89 (1999).

\bibitem{Sigl}
G.Sigl, F.Miniati, T.Ensslin, astro-ph/0401084

\bibitem{Hoerandel} J. R. Hoerandel, Astroparticle Phys. {\bf 19},
193 (2003);\\
%" Dissecting the knee - Air shower measurements with KASCADE"
Hoerandel J. R. et al (KASCADE Collaboration),  astro-ph/0311478.

\bibitem{Ptuskin}  V. S. Ptuskin et al, Astron. and Astroph.
{\bf 268}, 726 (1993).

\bibitem{Lagutin} A. A. Lagutin, Yu. A. Nikulin , V. V. Uchaikin, Nucl. Phys. B
(Proc. Suupl.), {\bf 97}, 267 (2001).

\bibitem{Roulet1}
%"Turbulent diffusion and drift in galactic magnetic fields and the explanation of the knee in the cosmic ray spectrum
J. Candia, E. Roulet, L. N. Epele,  JHEP 0212, 033 (2002).
% "Rigidity dependent knee and cosmic ray induced high energy neutrino fluxes"
%Julián Candia, Esteban Roulet, JCAP 0309 (2003) 005, astro-ph/0306632
%COSMIC RAY DRIFT, THE SECOND KNEE AND GALACTIC ANISOTROPIES.
%By Julian Candia (La Plata U.), Silvia Mollerach, Esteban Roulet (Buenos Aires, CONICET),. Jul 2002. 10pp.
%Published in JHEP 0212:032,2002
%astro-ph/0207143

\bibitem{Biermann} P. L. Biermann et al,\ 2003, astro-ph/0302201

\bibitem{Berezhko}
E. G. Berezhko,  L. T. Ksenofontov, JETP {\bf 89}, 391 (1999).

\bibitem{Erlykin} A. D. Erlykin and A. W. Wolfendale,
J. Phys. G.: Nucl. Part. Phys., {\bf 27}, 1005 (2001).

\bibitem{Tkachyk} W. Tkachyk, Proc. 27th Int. Cosmic Ray Conf.,
Hamburg, {\bf 5}, 1979 (2001);\\
S. Karacula  and W. Tkachyk, Astroparticle Phys., {\bf 1}, 229 (1993).

\bibitem{Roulet2}
J. Candia, L. N. Epele, E. Roulet, Astropart. Phys., {\bf 17}, 23 (2002).
%Title: Cosmic ray photodisintegration and the knee of the spectrum


\bibitem{kascade-g}  K.-H. Kampert et al. (the KASCADE-Grande Collaboration),
Proceedings of "XII International Symposium on Very High Energy Cosmic
Ray Interactions", CERN, Geneva, Switzerland 15-20 July 2002, astro-ph/0212347

\end{thebibliography}
\end{document}